\documentstyle[12pt,psfig]{article}
\begin{document}
\newcommand{\be}{\begin{equation}}
\newcommand{\ee}{\end{equation}}
\newcommand{\ba}{\begin{eqnarray}}
\newcommand{\ea}{\end{eqnarray}}
\newcommand{\bi}{\bibitem}
\def\Res{\mathop{\rm Res}\nolimits}
\def\Re{\mathop{\rm Re}\nolimits}
\def\Im{\mathop{\rm Im}\nolimits}
\large \bf
\begin{center}
Nonperturbative contributions in an analytic running coupling of QCD
\end{center}
\vskip 1cm
\normalsize
\begin{center}
 Aleksey I. Alekseev
\footnote{Electronic address: 
alekseev@mx.ihep.su}\\
{\small\it Institute for High Energy Physics,
142280 Protvino, Moscow Region, Russia}\\
\end{center}
\vskip 1.5cm
\begin{abstract}
In the  framework of analytic approach to QCD  the nonperturbative 
contributions in running coupling of strong interaction up to 4-loop
order  are obtained in an explicit form. For all $Q>\Lambda$ they are
shown to be represented in the form of
an expansion in  inverse powers of Euclidean momentum squared.
The expansion coefficients are calculated for different numbers
of active quark flavors $n_f$ and for different number of loops 
taken into account.  On basis of the stated expansion the effective
method for precise
calculation of the analytic running coupling can be developed. 
\end{abstract}
\vskip 1cm
PACS number(s): 12.38.Aw, 12.38.Lg
\vskip 1cm

It is widely believed that nonphysical singularities of the 
perturbation theory in the infrared region of QCD should be
canceled by the nonperturbative contributions.
The nonperturbative contributions arise quite naturally in an
analytic approach~\cite{SolShirTMF} --- \cite{Grunberg} to QCD.
The idea
of the approach goes back to Refs.~\cite{Red,Bog} devoted to the
nonphysical ghost pole problem in QED.   
In recent papers~\cite{Shir,Shir1} it is suggested to solve the ghost
pole problem in QCD  demanding the running coupling constant
be analytic in $Q^2$ ($Q^2$ is the Euclidean momentum squared). 
As a result of the procedure, instead of the one-loop expression 
$\alpha^{(1)}_s(Q^2)=(4\pi/b_0)/\ln (Q^2/\Lambda^2)$
taking into account the leading logarithms and having the ghost
pole at $Q^2=\Lambda^2$,
one obtains new  expression
\be
\alpha^{(1)}_{an}(Q^2)=\frac{4\pi}{b_0}\left[\frac{1}{\ln(Q^2/
\Lambda^2)}+\frac{\Lambda^2}{\Lambda^2-Q^2}\right].
\label{1}
\ee
Eq.~(\ref{1}) is an analytic function in the complex $Q^2$-plane
with a cut along the negative real semiaxis. The pole of the
perturbative running coupling at $Q^2=\Lambda^2$ is canceled by the
nonperturbative contribution 
$(\Lambda^2$ $\simeq \mu^2\exp
\{-4\pi/(b_0\alpha_{an}(\mu^2))\}$ at $\alpha_{an}(\mu^2)
\rightarrow
0$) and the value $\alpha^{(1)}_{an}(0)=
4\pi/b_0$ appeared to be finite and independent of the normalization
conditions (independent of $\Lambda$). 
The most important feature of the  procedure 
is the stability property~\cite{Shir,Shir1} of the value
of the 
"analytically improved" running coupling constant at zero with 
respect to high corrections, 
$\alpha^{(1)}_{an}(0)=$	$\alpha^{(2)}_{an}(0)=$
$\alpha^{(3)}_{an}(0)$.
This property provides the high corrections stability of 
$\alpha_{an}(Q^2)$
in the whole infrared region.
We shall demonstrate that $\alpha_{an}(0)=4\pi/b_0$ for any finite
loop order initial $\alpha_s(Q^2)$ in the standard inverse powers of
logarithms expansion form being a consequence of the asymptotic
freedom property.

Note that for the  one-loop case we have not only the spectral
representation for $\alpha_{an}(Q^2)$ but, in the first place, the
nonperturbative
contribution is extracted from the analytic running coupling
explicitly,
\be
\alpha_{an}(Q^2)=\alpha^{pt}(Q^2)+\alpha_{an}^{npt}(Q^2),
\label{m13}
\ee
and, in the second place, the 1-loop order nonperturbative
contribution
can be
presented 
as convergent at $Q^2>\Lambda^2$ of constant signs series in the
inverse
powers of the momentum squared,
\be
\alpha_{an}^{npt}(Q^2)=\frac{4\pi}{b_0}\sum\limits^{\infty}\limits_
{n=1}c_n\left(\frac{\Lambda^2}{Q^2}\right)^n,
\label{m14}
\ee
where $c_n=-1$. 
For the standard as well as for the iterative 2-loop perturbative
input
the nonperturbative contributions in the analytic
running coupling are calculated explicitly in Ref.~\cite{A}.
In the ultraviolet region the nonperturbative contributions also can 
be represented as a series in inverse powers of the momentum squared.
The 3-loop case is considered in Ref.~\cite{IHEP40} where, to handle
the singularities originating from the perturbative input, the method
which is more general then that of
Ref.~\cite{A} was developed.

In this paper we extract in an explicit form the nonperturbative 
contributions to $\alpha_{an}(Q^2)$	up to 4-loop order in
analytic approach to QCD, and find the coefficients $c_n$ of the
expansion of the form~(\ref{m14}). This gives the effective method
for a precise calculation of $\alpha_{an}(Q^2)$ which is not
connected with numerical integration.

The behavior of the QCD running coupling $\alpha_s(Q^2)$
is defined by the renormalization group equation
\be
Q^2\frac{\partial\alpha_s(Q^2)}{\partial Q^2}=
\beta(\alpha_s)=\beta_0\alpha_s^2+\beta_1\alpha_s^3+\beta_2\alpha_s^4
+\beta_3\alpha_s^5+O(\alpha_s^6),
\label{3}
\ee
where the coefficients~\cite{Gross} --- \cite{Larin97}
\ba
\beta_0=&-&\frac{1}{4\pi}b_0, \,\,\, b_0=11-\frac{2}{3}n_f,
\nonumber\\
\beta_1=&-&\frac{1}{8\pi^2}b_1, \,\,\, b_1=51-\frac{19}{3}n_f,
\nonumber\\
\beta_2=&-&\frac{1}{128\pi^3}b_2, \,\,\, b_2=
2857-\frac{5033}{9}n_f +\frac{325}{27}n_f^2,
\nonumber\\
\beta_3=&-&\frac{1}{256\pi^4}b_3, \,\,\, b_3=\frac{149753}{6}+
3564\zeta_3
\nonumber\\
&-&\left(\frac{1078361}{162}+\frac{6508}{27}\zeta_3\right)n_f+
\left(\frac{50065}{162}+\frac{6472}{81}\zeta_3\right)n_f^2+
\frac{1093}{729}n_f^3.
\label{4}
\ea
Here $n_f$ is number of active quark
flavors and $\zeta$ is Riemann zeta-function, $\zeta_3=\zeta(3)=
1.202056903...\,$. 
The first two coefficients $\beta_0$, $\beta_1$ do not depend on
the renormalization scheme choice. The next coefficients do depend on
this choice. Being calculated within the
$\overline{MS}$-scheme in an arbitrary covariant gauge for the gluon
field they appeared to be independent of the gauge parameter choice.
Integrating Eq.~(\ref{3}) yields
\ba
\frac{1}{\alpha_s(Q^2)}&+&\frac{\beta_1}{\beta_0}\ln\alpha_s(Q^2)+
\frac{1}{\beta_0^2}\left(\beta_0\beta_2-\beta_1^2\right)
\alpha_s(Q^2)+\frac{1}{2\beta_0^3}\left(\beta_1^3-
2\beta_0\beta_1\beta_2+\beta_0^2\beta_3\right)\alpha_s^2(Q^2)
\nonumber\\
&+&O(\alpha_s^3(Q^2))=-\beta_0\ln(Q^2/\Lambda^2)+\tilde{C}.
\label{5}
\ea
The integration constant is represented here as a combination
of two  constants $\Lambda$ and $\tilde{C}$.
Dimensional constant $\Lambda$ is a parameter which defines the 
scale of $Q$ and is used  for developing the iteration
procedure.
Iteratively solving Eq.~(\ref{5}) for $\alpha_s(Q^2)$ at
$L=\ln(Q^2/\Lambda^2)\rightarrow\infty$ one obtains
$$
\alpha_s(Q^2)=-\frac{1}{\beta_0 L}\left\{1+\frac{\beta_1}{\beta_0
^2L}\left(\ln L+C\right)+\frac{\beta_1^2}{\beta_0^4L^2}
\left[\left(\ln L+C\right)^2-\left(\ln L+C\right)-1+\frac{\beta_0
\beta_2}{\beta_1^2}\right]\right.
$$
\be
+\left.\frac{\beta_1^3}{\beta_0^6L^3}\left[\left(\ln L+C\right)^3
-\frac{5}{2}\left(\ln L+C\right)^2-\left(2-\frac{3\beta_0\beta_2}
{\beta_1^2}\right)\left(\ln L+C\right)+\frac{1}{2}-\frac{\beta_0^2
\beta_3}{2\beta_1^3}\right]+O\left(\frac{1}{L^4}\right)\right\},
\label{7}
\ee
where $C=\ln(-\beta_0)+(\beta_0/\beta_1)\tilde{C}$. Within the
conventional definition of $\Lambda$ as 
$\Lambda_{\overline{MS}}$ \cite{Bardeen78} one chooses 
$C=0$. At that the functional form of the approximate solution for
$\alpha_s(Q^2)$ turns out to be somewhat simpler, but it
requires distinct $\Lambda_{\overline{MS}}$ for different $n_f$.
With this choice Eq.~(\ref{7})  at three loop level corresponds to
the standard  solution  written in the form of the expansion 
in inverse powers of logarithms~\cite{Data},
and at four loop level it corresponds to~\cite{Chet97}.
We shall deal with nonzero $C$ because this freedom  can be useful
for an optimization of the finite order perturbation calculations.
Moreover, in the presence of the  $n_f$-dependent constant $C$ it is
possible to construct matched solution of Eq.~(\ref{3}) with
universal  $n_f$ independent constant $\Lambda$ \cite{Marciano}.

Let us introduce the function $\Phi(z)$ of the form
\ba
\Phi(z)&=&\frac{1}{z}-b\frac{\ln(z)+C}{z^2}
+b^2\left[\frac{\left(\ln(z)+C\right)^2}{z^3}-\frac{\ln(z)
+C}{z^3}
+\frac{\kappa}{z^3}\right]
\nonumber\\
&-&b^3\left[\frac{\left(\ln(z)+C\right)^3}{z^4 }-\frac{5}{2}
\frac{\left(\ln(z)+C\right)^2}{z^4}+(3\kappa+1)\frac{
\ln(z)+C}{z^4}+\frac{\bar\kappa}{z^4}\right],
\label{9}
\ea
where the coefficients $b$, $\kappa$, and $\bar\kappa$ are equal to
\ba
b&=&-\frac{\beta_1}{\beta^2_0}=\frac{2b_1}{b^2_0},\nonumber\\
\kappa&=&-1+\frac{\beta_0 \beta_2}{\beta_1^2}=-1+\frac{b_0 b_2}
{8b^2_1},\nonumber\\
\bar\kappa&=&\frac{1}{2}-\frac{\beta_0^2\beta_3}{2\beta_1^3}=
\frac{1}{2}-\frac{b_0^2b_3}{16b_1^3}.
\label{10}
\ea
To choose the main branch of the multivalued function~(\ref{9}) we 
cut complex z-plane along the negative semiaxis. Then the
solution~(\ref{7}) can be written as 
$\alpha_s(Q^2)=(4\pi/b_0)a(x)$, where $a(x)=\Phi(\ln x)$,
$x=Q^2/\Lambda^2$.  Function $a(x)$ is unambiguously defined in the
complex $x$-plain with two cuts along the real axis, physical cut
from minus infinity to zero and nonphysical one from zero to unity. 
At $x\simeq 1$   the perturbative running coupling has singularities
of a different analytical structure.
Namely, at $x\simeq 1$ the leading singularities are
\ba
a^{(1)}(x)&\simeq&\frac{1}{x-1},\,\,\,
a^{(2)}(x)\simeq-\frac{b}{(x-1)^2}\ln (x-1),\nonumber\\
a^{(3)}(x)&\simeq&\frac{b^2}{(x-1)^3}\ln^2 (x-1),\,\,\,
a^{(4)}(x)\simeq-\frac{b^3}{(x-1)^4}\ln^3 (x-1).
\label{11}
\ea
This is not an obstacle for the analytic approach which removes
all this nonphysical singularities.

The analytic running coupling is defined  by the spectral
representation
\be
a_{an}(x)=\frac{1}{\pi}\int\limits_0^\infty \frac{d\sigma}{x+\sigma}
\rho(\sigma),
\label{2}
\ee
with the spectral density $\rho(\sigma)=\Im a_{an}(-\sigma-i0)=\Im
a(-\sigma-i0)$ where $a(x)$ is
the  perturbative running coupling.
It is seen that dispersively modified coupling of the 
form~(\ref{2}) has
analytical structure which is consistent with causality.

By making the analytic continuation of  Eq.~(\ref{9}) into 
the Minkowski
space by formal substitution $x=-\sigma-i0$  one  obtains
the spectral density as an imaginary part of $\Phi(\ln
\sigma-i\pi)$.
Function $a(x)$ is regular and real for real $x>1$.
So, to find the spectral density $\rho(\sigma)$ we can use the 
reflection principle
$(a(x))^*=a(x^*)$ where $x$ is considered as a complex
variable. Then
\be
\rho(\sigma)=\frac{1}{2i}\left( \Phi(\ln\sigma-i\pi)-\Phi(\ln
\sigma+i\pi)\right).
\label{13}
\ee
By the change of  variable of the form $\sigma=\exp (t)$,
the analytical expression
is derived from~(\ref{2}),  (\ref{13}) as follows:
\be
a_{an}(x)=\frac{1}{2\pi i}\int\limits^\infty_{-\infty}
dt \, \frac{e^t}{x+e^t}\times
\left\{\Phi(t-i\pi)-\Phi(t+i\pi)
\right\}.
\label{14}
\ee
Let us prove that $a_{an}(0)=1$. It follows from Eq.~(\ref{14}) that
$$
a_{an}(0)=\frac{1}{2\pi i}\int\limits^\infty_{-\infty}
dt \, 
\left\{\Phi(t-i\pi)-\Phi(t+i\pi)
\right\}
$$
\be
=\frac{1}{2\pi i}\int\limits^\infty_{-\infty}
dt \, 
\left\{\left[\Phi(t-i\pi)-\frac{1}{t-i\pi}\right]-\left[\Phi(t+i\pi)-
\frac{1}{t+i\pi}\right]+\left[\frac{1}{t-i\pi}-\frac{1}{t+i\pi}\right
]
\right\}.
\label{144}
\ee
For the first term in Eq.~(\ref{144}) close the integration contour
in the lower half-plane of the complex variable $t$ by the arch of
the "infinite" radius without 
affecting the value of the integral. We can do it because
the integrand  multiplied by $t$ goes to
zero at $\mid t\mid\rightarrow \infty$.
There are no singularities
inside the contour, so we obtain zero contribution from the term
considered. For the second term  we close the integration contour
in the upper half-plane of the complex variable $t$ with the same
result. So we have
\be
a_{an}(0)=\frac{1}{2\pi i}\int\limits^\infty_{-\infty}
dt \, 
\left[\frac{1}{t-i\pi}-\frac{1}{t+i\pi}\right]=1.
\label{145}
\ee
For any finite loop order the expansion structure of the perturbative
solution in inverse powers of logarithms ensure the property of the
analytical coupling $a_{an}(0)=1$. The arguments are suitable for all
solutions $\Phi(z)$ if singularities are situated at the real axis of
complex $z$-plane in particular for the iterative solutions of
Refs.~\cite{Shir,Shir1}.

Let us see what the singularities of the integrand of~(\ref{14}) 
in the complex  $t$-plane are.
First of all the integrand has  simple
poles at  $t=\ln x\pm i\pi(1+2n)$,
$n=0,1,2,...$. All the residues of the function  
$\exp(t)/(x+\exp(t))$ 
at these points are equal to unity.
Apart from this 
poles the integrand of~(\ref{14}) 
has at $t=\pm i\pi$  poles up to fourth order  and
logarithmic type branch points  which coincide with poles
from  second order to fourth order. Let us cut the complex
$t$-plane in a  
standard way, $t=\pm i\pi-\lambda$, with $\lambda$ being the real
parameter varying from  $0$ to $\infty$.
Append
the integration by the arch of the "infinite" radius without 
affecting the value of the integral. Close the integration contour
$C_1$ in the upper half-plane of the complex variable $t$
excluding the singularities at $t=i\pi$. In this case 
an additional contribution emerges due to the integration along
the sides of the cut and around the singularities at $t=i\pi$.
The corresponding contour we denote as $C_2$.

\begin{figure}[tbp]
\centerline{\psfig{figure=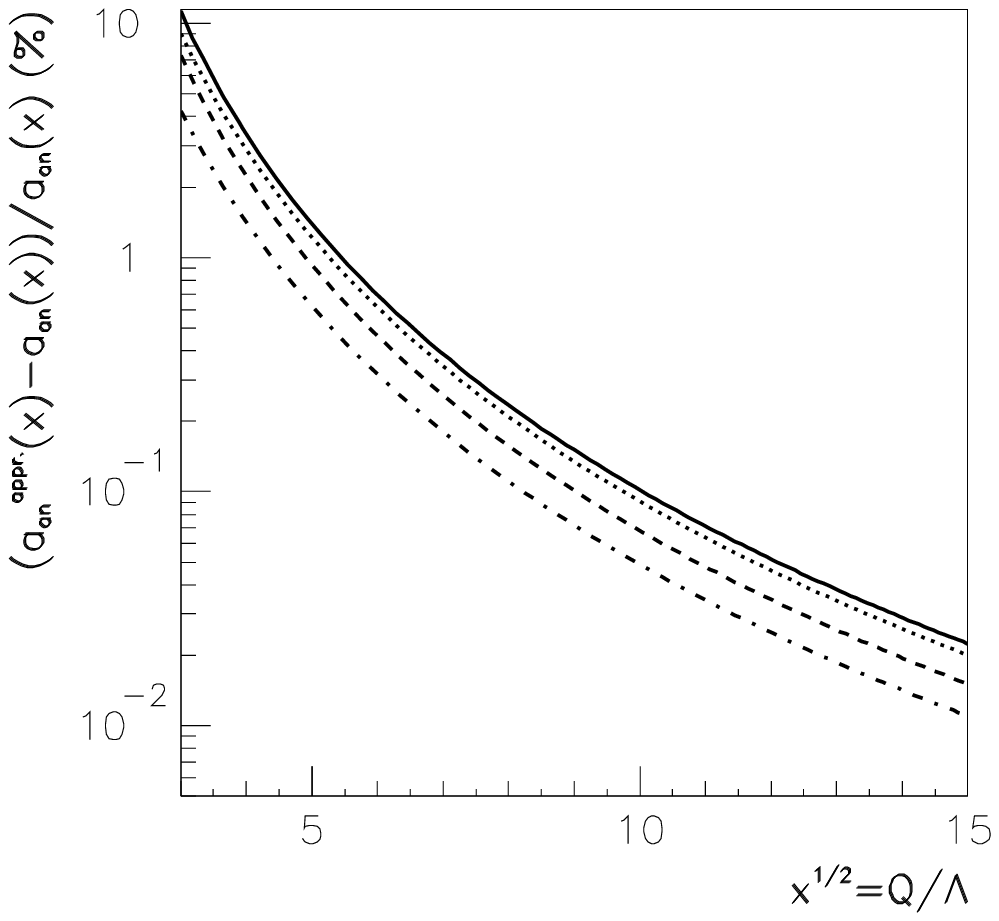,height=12cm,width=12cm}}
\caption{Relative error of the approximation of $a_{an}$ 
with $a_{an}^{npt}$ approximated by one first term of the series,  
as function of
$x^{1/2}=Q/\Lambda$ for 1 --- 4-loop cases  at $n_f=4$.
Dash-dotted line, dotted line, dashed line and solid line correspond
to 1-loop, 2-loop, 3-loop and 4-loop cases respectively.
}
\label{fig5}
\end{figure}

Let us turn to the integration along the contour $C_1$.
For the integrand of Eq.~(\ref{14}) which we denote as
$F(t)$ the residues at $t=\ln x+i\pi(1+2n)$, $n=0,1,2,...$  are
as follows
\be
\Res F(t)\mid _{t=\ln x+i\pi(1+2n)}=
\Phi(\ln x+2\pi i n)-\Phi(\ln x+2\pi i(n+1)).
\label{200}
\ee
By using the residue theorem  
one readily obtains  the  contribution $\Delta(x)$ to the
integral~(\ref{14}) from the integration along contour $C_1$.
It reads
\be
\Delta(x)=\frac{1}{2\pi i}\int\limits_{C_1}F(t)\,dt= 
\sum\limits^\infty_{n=0}\Res F\left(t=\ln x+i\pi(1+2n)\right)
=\Phi(\ln x).
\label{16}
\ee
We can see that this contribution is exactly equal to the initial
$a(x)$. Therefore we call it a perturbative part of 
$a_{an}(x)$,  $a^{pt}(x)=\Delta(x)$.
The remaining contribution of the integral along the contour $C_2$
can naturally be called a nonperturbative part of $a_{an}(x)$
according to Eq.~(\ref{m13}).
Let us turn to the calculation of  $a^{npt}_{an}(x)$.  We have
\be
a^{npt}_{an}(x)=\frac{1}{2\pi i}\int\limits_{C_2}
dt \, \frac{e^t}{x+e^t}\times
\left\{\Phi(t-i\pi)-\Phi(t+i\pi)\right\}.
\label{18}
\ee
We consider $x$   as real  variable, $x>1$. One can omit the
term of the integrand in Eq.~(\ref{18}) which has 
singularities at $t=-i\pi$. Let us change the variable $t=z+i\pi$ and
introduce the function
\be
f(z)=\frac{1}{1-x\exp(-z)}.
\label{19}
\ee
Then we can rewrite Eq.~(\ref{18}) in the form
$$
a^{npt}_{an}(x)=\frac{1}{2\pi i}\int\limits_{C} dz \, f(z)
\left\{\frac{1}{z}-b\left[\frac{\ln(z)}{z^2}+\frac{C}{z^2}\right]
+b^2\left[\frac{\ln^2(z)}{z^3}+\left(2C-1\right)\frac{\ln z}{z^3}+
\frac{\kappa-C+C^2}{z^3}\right]\right.
$$
\be
-\left.b^3\left[\frac{\ln^3(z)}{z^4}+\left(3C-\frac{5}{2}\right)
\frac{\ln^2}{z^4}+(3C^2-5C+3\kappa+1)\frac{\ln z}{z^4}+
\frac{C^3-\frac{5}{2}C^2+(3\kappa+1)C+\bar\kappa}{z^4}\right]
\right\}.
\label{20}
\ee
The cut in  the complex $z$-plane goes now from zero to $-\infty$.
Starting from $z=-\infty-i0$ the contour $C$ goes 
along the lower side of the cut then
goes around the origin, 
and then it goes further along the upper side of the cut  to  
$z=-\infty+i0$. 
The contour $C$ can be chosen in such a way that it does not
envelop a "superfluous" singularities corresponding to the
perturbative contributions. 
Using the technique of Ref.~\cite{IHEP40} one can integrate the terms
of Eq.~(\ref{20}) and obtain
$$
a^{npt}_{an}(x)=-\frac{1}{x-1}+b\left\{\frac{(1+C)x}{(x-1)^2}+
x\int\limits_{0}\limits^{1} d\sigma \,\ln(-\ln\sigma)
\frac{x+\sigma}{(x-\sigma)^3}\right\}
$$
$$
-\frac{1}{2}b^2\left\{\left[1-\frac{\pi^2}{3}+\kappa+(1+C)^2\right]
\frac{x(x+1)}{(x-1)^3}
+x\int\limits_{0}\limits^{1} d\sigma \,\biggl[
2(1+C)\ln(-\ln\sigma)+\ln^2(-\ln\sigma)\biggr]\right.
$$
$$
\times\left.
\frac{x^2+4x\sigma+
\sigma^2}{(x-\sigma)^4}\right\}
+\frac{1}{6}b^3\left\{\biggl[2+\frac{5}{2}\kappa  
+\bar\kappa+3(1+C)
\left(1-\frac{\pi^2}{3}+\kappa\right) +(1+C)^3\biggr]\right.
$$
$$
\times\frac{x(x^2+4x+1)}{(x-1)^4}
+x\int\limits_{0}\limits^{1} d\sigma \,\biggl[
3\biggl(1-\frac{\pi^2}{3}+\kappa+(1+C)^2\biggr)	
$$
\be
\times
\ln(-\ln\sigma)
+3(1+C)\ln^2(-\ln\sigma)+\ln^3(-\ln\sigma)
\biggr]\frac{x^3+11x^2\sigma+11x\sigma^2+\sigma^3}{(x-\sigma)^5}
\Biggr\}.
\label{26}
\ee
This formula gives the nonperturbative contributions in the explicit
form.
Expanding Eq.~(\ref{26}) in inverse powers of $x$ 
we have
\be
a_{an}^{npt}(x)=\sum\limits_{n=1}\limits^{\infty}\frac{c_n}{x^n}.
\label{28}
\ee
Making the change of variable $\sigma=\exp(-t)$ and integrating 
\cite{Bateman}, \cite{Prudnikov} over $t$  we  finally  find
$$
c_n=-1+bn\left[1+C-\gamma-\ln(n)\right]-\frac{1}{2}b^2n^2\left[
1-\frac{\pi^2}{6}+\kappa+\Bigl(1+C-\gamma-\ln(n)\Bigr)^2\right]
$$
\be
+\frac{1}{6}b^3n^3\left[2+\frac{5}{2}\kappa+\bar\kappa-
2\zeta_3+\Bigl(1+C-\gamma-\ln (n)\Bigr)^3+
3\Bigl(1+C-\gamma-\ln (n)\Bigr)\left(1-\frac{\pi^2}{6}+\kappa\right)
\right].
\label{31}
\ee
Here  $\gamma$ is the Euler constant,  $\gamma\simeq 0.5772$.
We can see from Eq.~(\ref{31})
that power series~(\ref{28}) is uniformly convergent at $x>1$
and its convergence radius is  equal to unity.
For numerical evaluation of the coefficients $c_n$
assume that $C=0$. Then the coefficients $c_n$ depend on $n$, $n_f$
and number of loops
taken into account. The 1-loop order contribution to $c_n$ equals
to $-1$ for all $n$ and $n_f$. Up to 4-loop approximation the
coefficients $c_n$ for all  $n$, $n_f$ are negative. 
With the exception of the 3-loop case at $n_f=6$,
the 2 --- 4-loop
coefficients $c_n$ for $n_f=0,3,4,5,6$ monotonously increases in the
absolute value with increasing $n$.
In the ultraviolet region ($x\gg 1$) the nonperturbative
contributions 
are determined by the first term of the series~(\ref{28}).
In Table~\ref{t2} we give the values of $c_1$ and loop corrections 
for $n_f=0,3,4,5,6$.
\begin{table}[tbp]
\caption{The dependence of $c_1$ on  $n_f$ for 1-loop, 2-loop, 
3-loop, and 4-loop cases.}
\label{t2}
\begin{center}
\begin{tabular}{ r r r r r r r r}\hline \hline
     $n_f$& $c^{1-loop}_1$  &$\Delta_{2-loop}$  & $\Delta_{3-loop}$ 
	 & $\Delta_{4-loop}$& $c_1^{2-loop}$ &$c_1^{3-loop}$ &
	 $c_1^{4-loop}$\\ \hline

0 &-1.0
&   0.35640&  -0.01568&  -0.03900&  -0.64360&  -0.65929&  -0.69828\\
	 \hline
3&-1.0
&   0.33405&   0.01608&  -0.07825&  -0.66595&  -0.64987&  -0.72812\\
	 \hline
4&-1.0
&   0.31252&   0.04949&  -0.11006&  -0.68748&  -0.63799&  -0.74805\\
	 \hline
5&-1.0
&   0.27813&   0.11653&  -0.16002&  -0.72187&  -0.60535&  -0.76537\\
	 \hline
6&-1.0
&   0.22433&   0.25378&  -0.22731&  -0.77567&  -0.52189&  -0.74920\\
	 \hline \hline

\end{tabular}
\end{center}
\end{table}
One can  see that for all $n_f$ up to four loops $c_1$ is 
of the order of unity.
An account for the high loop corrections results in some 
compensation of the 1-loop  leading at large $x$ term of the
form $1/x$.
The relative error of the approximation of the analytic running
coupling
with only one first term of the series~(\ref{28}) for the
nonperturbative
contributions taken into account,
$$
\alpha_{an}(Q^2)\simeq\alpha^{pt}(Q^2)-\frac{4\pi}{b_0}
\left\{1-b\left(1-\gamma\right)+\frac{1}{2}b^2\left(
1-\frac{\pi^2}{6}+\kappa+\left(1-\gamma\right)^2\right)\right.
$$
\be
\left.
-\frac{1}{6}b^3\left[2+\frac{2}{5}\kappa+\bar\kappa-2\zeta_3+
\left(1-\gamma\right)^3+3\left(1-\gamma\right)\left(1-\frac{\pi^2}{6}
+\kappa\right)\right]
\right\}
\frac{\Lambda^2}{Q^2},
\label{n3}
\ee
is shown in Fig.~\ref{fig5} for 1 --- 4-loop cases.
We have chosen $n_f=4$ within  the region of $Q$ considered.

As a result of the expansion  coefficients increase not too fast,
the representation of the analytic running coupling of QCD
in the form~(\ref{m13}), (\ref{m14}) with $c_n$ as~(\ref{31}) 
provide one with the effective method for the calculation of 
$\alpha_{an}$ at $Q>\Lambda$.  At that there is no need for the
summation
of large number of terms of the series. 
Approximation of the nonperturbative "tail" by the leading term
already at $Q\sim 5\Lambda$ gives one percent accuracy for
$\alpha_{an}$. As regards the parameters $\Lambda^{(n_f=5)}$,
$\Lambda^{(n_f=4)}$ 
of the analytic approach they turn out to be close to the
perturbative 
values as a consequence of the rapid decrease of the nonperturbative
contributions in the ultraviolet region.

I am indebted to B.A.~Arbuzov, V.A.~Petrov, V.E.~Rochev for 
useful discussions. This work was  supported in part by the 
Russian Foundation for Basic Research under
grant No.~99-01-00091.

%

\end{document}